# TOWARDS A MOLECULAR INVENTORY OF PROTOSTELLAR DISCS

**Glenn J White** [1], **Mark A. Thompson** [1], **C.V. Malcolm Fridlund** [2], **Monica Huldtgren White** [3]

[1] University of Kent, UK, g.j.white@kent.ac.uk  [2] ESTEC, NL,  [3] Stockholm Observatory, Sweden

**Abstract**

The chemical environment in circumstellar discs is a unique diagnostic of the thermal, physical and chemical environment. In this paper we examine the structure of star formation regions giving rise to low mass stars, and the chemical environment inside them, and the circumstellar discs around the developing stars.

**Introduction**

Life develops from a sequence of organic chemical processes that develop sufficient complexity to lead to mutating and self-replicating sentient organisms. Cellular structures, formed following selective evolution along the Tree of Life, emerged from the reactions amongst these simple organic compounds, which were derived from cosmically abundant elements and molecules that are abundant in space. Much of this early chemical inventory was delivered to the surfaces of planets, including the Earth, during the early phase of massive cometary bombardment of the earth - e.g. mixtures of methane, ammonia may have been amongst the precursors that led to the amino acids of proteins and the bases of nucleic acids in the RNA and DNA chains. To date, more than 20 molecules have been discovered in comets - their icy compositions are analogous to that of interstellar material in protostellar cores, suggesting that comets are relics of the protosolar chemical evolution of the natal star forming material. Although complex organics such as HCOOH or $HCOOCH_3$ are readily observed, the chemical inventory of material in protostellar and protoplanetary discs is far from complete. In the current absence of a ready source of pristine proto-nebular material, molecular line surveys from ground and space-based telescopes provide our best, and most unbiased view of the chemical inventory of material that must have existed in the early solar system; and the reaction pathways that trace the protoplanetary environment. As the resolution and fidelity of the next range of coronographic and interferometric imagers allow us to see exozodaical discs, study of the molecular and dusty components will allow studies of the incorporation of biogenic precursors in the natal clouds into the planet forming discs around stars.

In this paper, we report preliminary results from an on-going programme to search for triggered star formation regions – to understand the formation of stellar discs; and a molecular line survey towards protostellar discs to characterise the make-up of the solid, atomic and ionised material in the pre-planetary environment.

**Details of the survey**

We are carrying out a survey of nearby star formation regions to

- Search in the optical/IR region for triggered/induced solar-type star formation. We measure the boundary conditions, such as external pressure of ionised gas that triggering the star formation using radio observations, and from molecular line tracers, estimate the internal pressure in the molecular gas.
- Characterise the properties of the central protostellar core and protostellar discs at submm, mid and near-IR wavelengths, by measuring the spectral energy distributions and the atomic and molecular inventory of a sample of circumstellar discs
- Measure the chemical inventory of gas-phase material in the clouds using molecular line spectral surveys to characterise the protoplanetary environment.

**Bright Rim Cloud Survey**

Star formation is a complex process that forms stars ranging from ~ 0.1 – 60 $M_\odot$ - the present survey is aimed at identification of environments where solar mass protostellar objects may form. Although low mass star formation is widespread throughout dark clouds in the Galaxy, the exact details of the formation remain veiled due in part to the complexity of the natal environment. We report observations of a class of objects known as 'Bright Rim clouds' – these are traced by narrow ionised rims at the edges of quiescent dark clouds, illuminated by the UV radiation of nearby OB stars. An overpressure builds up on the ionised outer layers of the cloud – driving a shock wave into the dark cloud and establishing a photo evaporated flow that streams off the ionised skin of the cloud. This overpressure induces star formation inside the cloud within a few sound crossing times (~ several x $10^5$ years). Such objects have a set of boundary conditions: external pressure, temperature, density, moderated solely by their internal and external environments – and their properties can be measured with some accuracy. A well known example of a 'Bright Rim' cloud is the Eagle Nebula (Hester *et al* 1996, AJ, 111, 2349, White *et al* 1999, A&A, 342, 233). Solar mass stars are being forced to form inside the tips of each of the

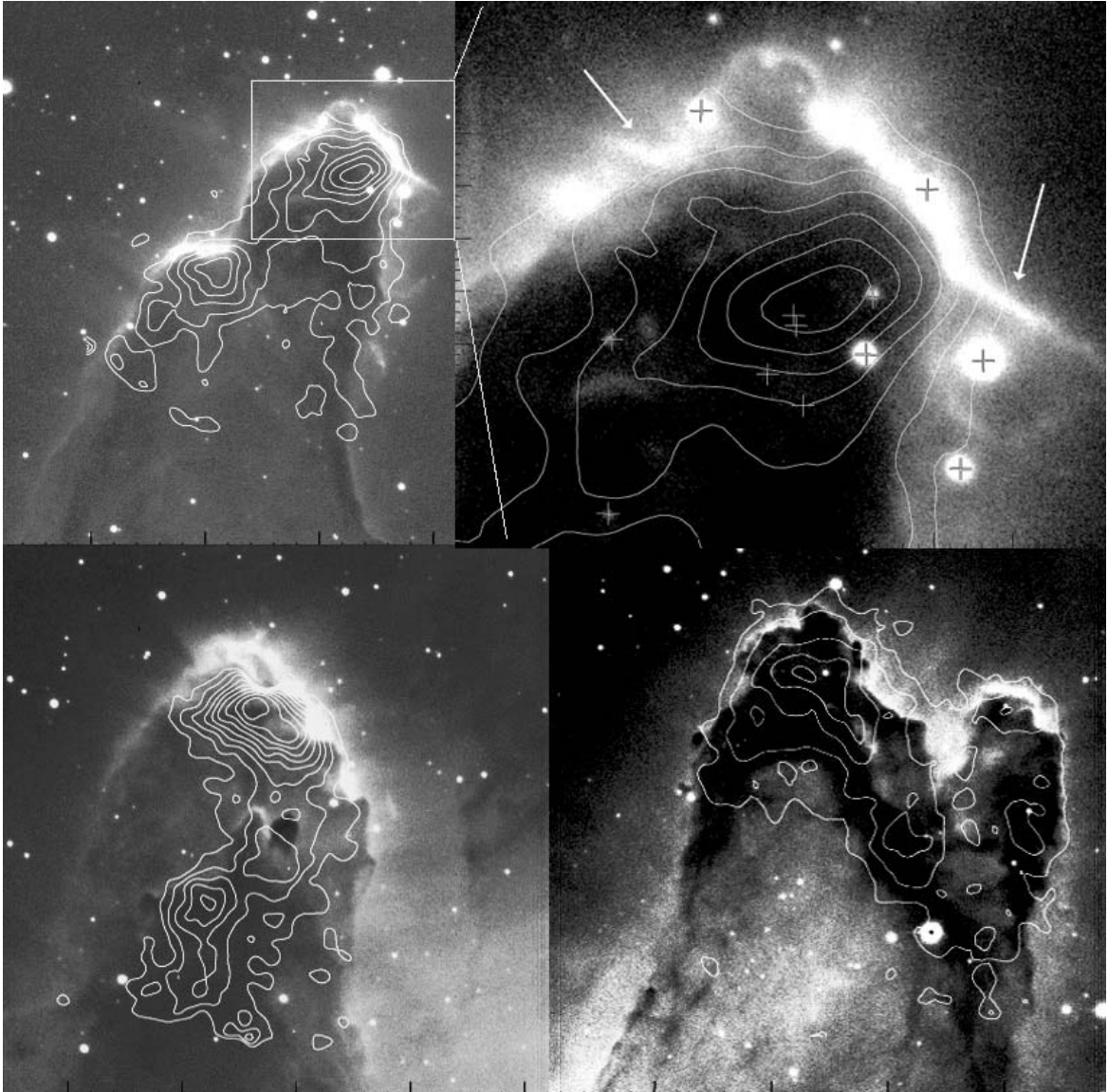

*Figure 1. Narrow band Hα images obtained with the Nordic Optical Telescope; contours show the 850 μm submm continuum obtained with SCUBA on the JCMT superposed. The tick marks are at 30 arc second intervals.*

Eagle Nebula's fingers, coevally with lower mass star, in corrugations at the cloud edges (McCaughrean & Andersen 2002). Examples of Bright Rim clouds are shown in Figure 1.

**The gas phase chemical inventory**

By observing at mm and submm wavelengths through the atmospheric 'windows', it is possible to construct the chemical inventory of the gas phase material out of which the protostars and the protoplanetary discs will form. A spectrum towards a 'hot core' in the heart of one of the star forming regions in Orion is shown in Figure 2.

**The protostellar chemical inventory**

Table 1 lists column densities of molecules observed between 462 – 506 GHz towards the Orion Nebula protistar. This gas-phase material is the reservoir of material available for incorporation into planetary disc(s). Theoretical studies of the chemistry of pre-planetary discs have been presented by Aikawa *et al* (1997) *et loc cite*. Although these models show that the chemical make up of simple ices, such as those which may be incorporated into cometary hosts, follow naturally from grain-surface reactions, a wider range of pre-cursor molecules must have been incorporated into the embryonic protostellar nebula to account for the abundances of complex organic molecules in comets, meteorites, and interstellar dust grains. An example of this is shown in Figure 3.

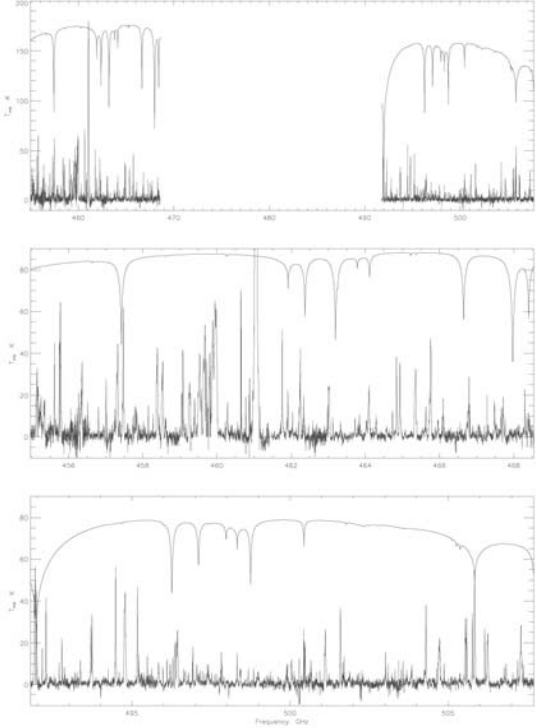

*Figure 2  462 – 506 GHz spectrum of a protostellar core in Orion showing 312 spectral lines revealing the complex chemistry in gaseous material (White et al 2003). The spectrum shows the remarkable chemical complexity that can be formed after a few million years of chemical evolution. Many of these species will eventually become incorporated into the pre-planetary nebula around individual star*

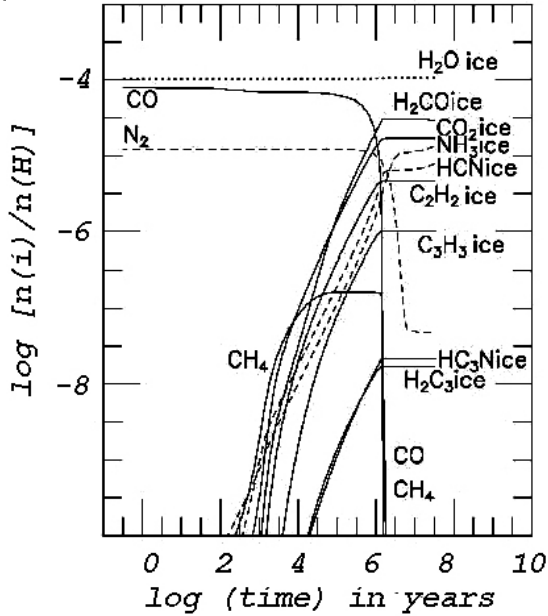

*Figure 3. Time evolution of simple ices and species in the proto-nebular environment, showing how simple species, initially present in the envelopes of accreting protostellar clouds, become incorporated into simple ices (adapted from Aikawa et al 1997)*

### The protostellar chemical inventory

Table 1 below lists the column densities of key lines observed in a small part of the submm wave spectrum (462 – 506 GHz) towards the protostellar region at the heart of the Orion Nebula. This represents primarily gas-phase material and forms the reservoir of material available for incorporation into a future planetary disc.

| Molecule | $N_{col}$ cm$^{-3}$ | Error cm$^{-3}$ | $T_{rot}$ K | Error K | Number of lines | Note |
|---|---|---|---|---|---|---|
| (CH$_3$)$_2$O | 1.4 10$^{16}$ | 1.8 10$^{15}$ | 157 | 30 | 27(26) lines | |
| C$_2$H$_3$CN | 3.0 10$^{17}$ | 3.2 10$^{17}$ | 180 | 47 | 13(6) lines | [1] |
| C$_2$H$_5$CN | 2.4 10$^{16}$ | 8.4 10$^{15}$ | 150 | 12 | 27(23) lines | $T_{rot}$ from Sut85 |
| | 8.3 10$^{15}$ | 1.2 10$^{15}$ | 239 | 12 | 27(23) lines | $T_{rot}$ from Sch01 |
| C$_2$H$_5$OH | 5.6 10$^{16}$ | 4.0 10$^{16}$ | 70 | - | 8(6) lines | $T_{rot}$ from Ohi95 |
| | 4.1 10$^{16}$ | 2.3 10$^{16}$ | 264 | 196 | 8(6) lines | |
| CH$_2$NH | 2.4 10$^{15}$ | 5.9 10$^{14}$ | 150 | - | 3(2) lines | $T_{rot}$ from HNCO |
| CH$_3$CN | 3.6 10$^{15}$ | 4.2 10$^{14}$ | 227 | 21 | 19 lines | |
| CH$_3$$^{13}$CN | 6.0 10$^{14}$ | 2.2 10$^{14}$ | 227 | - | 2 lines | $T_{rot}$ From CH$_3$CN |
| | 2.9 10$^{15}$ | - | 74 | - | 2 lines | |
| CH$_3$OH | 9.3 10$^{16}$ | 4.8 10$^{16}$ | 599 | 295 | 24(23) lines | |
| H$_2$CO | 1.6 10$^{16}$ | - | 166 | - | 2(1) lines | $T_{rot}$ from Bla87 |
| H$_2$$^{13}$CO | 1.0 10$^{15}$ | 4.0 10$^{14}$ | 166 | - | 2 lines | $T_{rot}$ from Bla87 |
| HC$_3$N | 1.5 10$^{15}$ | - | 164 | - | 2 lines | |
| HCOOCH$_3$ | 5.1 10$^{16}$ | 9.5 10$^{15}$ | 301 | 95 | 26(24) lines | |
| HNCO | 4.9 10$^{15}$ | 4.0 10$^{14}$ | 150 | 14 | 4(3) lines | |
| NH$_2$CN | 3.3 10$^{15}$ | 8.5 10$^{14}$ | 200 | - | 3 lines | $T_{rot}$ = 200 (K) Fix |
| | 1.1 10$^{16}$ | 4.8 10$^{15}$ | 100 | - | 3 lines | $T_{rot}$ = 100 (K) Fix |
| OCS | 9.0 10$^{16}$ | - | 106 | - | 2 lines | |
| SO | 3.3 10$^{17}$ | - | 72 | - | 2(1) lines | $T_{rot}$ from Sut95 |
| $^{34}$SO | 1.1 10$^{16}$ | 3.5 10$^{15}$ | 89 | 43 | 5(4) lines | |
| SO$_2$ | 1.2 10$^{17}$ | 1.0 10$^{16}$ | 136 | 9 | 35(28) lines | |
| $^{34}$SO$_2$ | 8.5 10$^{15}$ | - | 156 | - | 3(2) lines | |
| $^{13}$CS | 2.3 10$^{14}$ | - | 120 | - | 1 line | $T_{rot}$ from Zen95 |
| $^{30}$SiO | 3.4 10$^{14}$ | - | 50 | - | 1 line | $T_{rot}$ from Sut95 |
| CH$_3$CHO | 1.1 10$^{16}$ | - | 81 | - | 1 line | $T_{rot}$ from Sch97 |
| CI | 1.2 10$^{18}$ | - | 30 | - | 1 line | $T_{ex}$ from Whi95 |
| CO | 3.5 10$^{18}$ | - | 200 | - | 4 lines (1 transition) | $T_{rot}$ from Sut95 |
| DCN | 1.1 10$^{14}$ | - | 200 | - | 1 line | $T_{rot}$ from Bla87 |
| HCOOH | 2.2 10$^{15}$ | - | 100 | - | 1 line | $T_{rot}$ from Sut95 |
| HDO | 3.2 10$^{16}$ | - | 164 | - | 1 line | $T_{rot}$ from Bla87 |
| N$_2$O | 4.6 10$^{16}$ | - | 230 | - | 1 line | $T_{rot}$ from Wri83 |
| NH$_2$CHO | 7.5 10$^{15}$ | - | 81 | - | 1 lines | $T_{rot}$ from Sch97 |
| NH$_2$D | 8.7 10$^{15}$ | - | 160 | - | 1 line | $T_{rot}$ from Her88 |

*Table 1. Column densities, $N_{col}$, and rotational temperatures, $T_{rot}$, of molecular gas in one of the Orion protostellar cores.*

### The chemical inventory of Circumstellar Discs

The probability of significant chemical processing occurring as a result of two-body reactions in the gas phase is small, because of the low probability of collisions. Chemical reactions occur more efficiently on the surfaces of small dust grains, onto which molecular gas accretes, depleting gas in the envelope and disc of star-forming cores.. Observations at near, mid and far-IR wavelengths have been successful at detecting the chemical inventory of the gas in the solid and gaseous material associated with protostellar discs, much of which is incorporated into icy material that may be the precursor of the embryonic planetary systems. Models of disc chemistry have been presented by Aikawa *et al* (1997) and Markwick *et al* (2002). The spectra shown below in Figure 4 are taken from observations of the Lynds 1551 protostellar disc, reveal the emission lines from solid state and icy components, as well as that from ionised atomic and molecular gas.

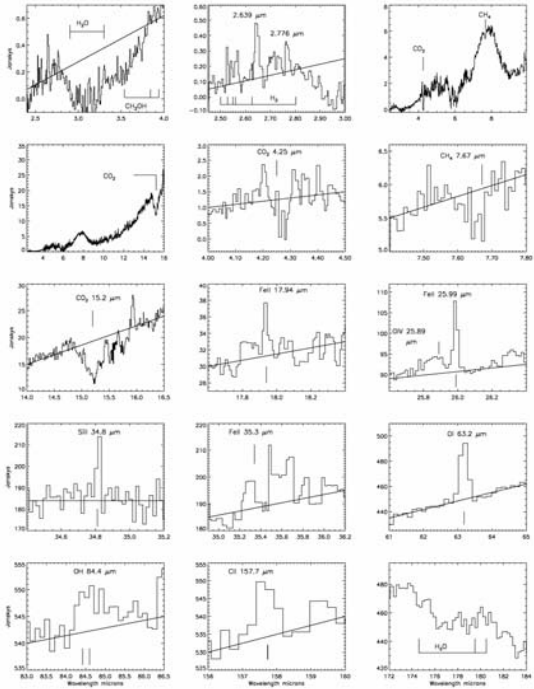

*Figure 4. ISO Long wavelength spectrometer spectra of the L1551 circumstellar disc (White et al 2000) showing icy features including $H_2O$, $CO_2$, $CH_4$, as well as atomic and molecular gas. This provides a clear view of the chemistry in the region (White et al 1999)*

**Protoplanetary discs**

Studies of the molecular inventories of the pre- and post circumstellar disc phase are important because they directly show the material that is available for incorporation into planetary bodies. Amongst the most successful techniques for studies of this material has been the use of mid- and far-infrared spectroscopy. This probes both the molecular, ionic and atomic gas, and can reveal the diversity of excitation that may be present. An example of a well studied circumstellar disc is that surrounding the binary protostellar system at the centre of the molecular outflow source associated with Lynds 1551. Our observations (Fridlund et al 2002) demonstrate the presence of a large ~ 7000 AU radius, dense, possibly rotating, molecular disk with a mass of a few $M_\odot$ oriented perpendicular to the major axis of an extended molecular outflow. The disk is surrounded by an envelope with a radius of ~ 10 000 AU that contains two massive (each ~ 1 $M_\odot$) clumps

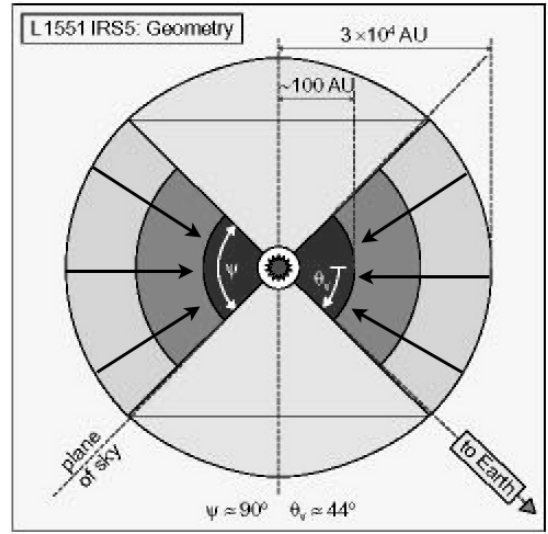

*Figure 5. Geometry of the protostellar core in L1551 IRS 5. Schematically shown are three regions of the model – the innermost dense torus (dark color), the constant-density part of the envelope (medium color), and the outer extended envelope containing most of the mass (light color). The bipolar geometry is defined by the opening angle of the conical outflow cavities ~ 90 degrees and the viewing angle, ~ 44 degrees to the line of sight (Figure adapted from White et al 2000).*

**Conclusions**

Studies of the molecular inventories of the pre- and post circumstellar disc phase are important because they directly show what material the bodies in the planetary systems are made from, and the molecular abundances are a useful probe in investigating the formation processes of planetary systems and primitive bodies, such as comets. Studies of the structure and evolution of the disk by observations of molecular lines also constrain the amount of the gaseous component in the disk, particularly CO and $N_2$ which are amongst the dominant components in the protoplanetary nebula. This allows us to probe the evolution from massive gas-rich disks with masses ~ 0.01 $M_\odot$ and lifetimes ~ 1 Myr to the more tenuous debris-discs within which primitive bodies and planets will condense. Key questions that will be solved include:

- The physical structures of the discs – $T$, $\rho$, $n$, $v$
- Chemical evolution of the dust and gaseous components
- Timescales for dust and gas dissipation and cycling into primitive bodies
- Evidence for planet formation

Surveys such as the one outlined in this paper are an important probe of the environment in regions where planets may be emerging from their placental circumstellar discs. The next generation of large submm, far and mid infrared telescopes, including ASTRO-F, ALMA, Herschel and the VLTI have the potential to probe this environment with a major step forward in sensitivity and angular resolution during the next 5 – 10 years. DARWIN itself then will have the possibility to detect the atmospheric environment of individual Exoplanets, allowing us to couple these to the molecular inventories of the material out of which they formed.